\pdfoutput=1

\documentclass[12pt]{article}

\usepackage{graphicx}
\usepackage[varg]{txfonts} % Times fonts
\usepackage[latin1]{inputenc}
\usepackage{hyperref}

\usepackage{slashbox,hhline}
\usepackage{xcolor}
\usepackage{multicol}

\usepackage{bm}
\newcommand{\etal}{\MakeLowercase{\textit{et al.}}}
\hypersetup{%
    pdftitle = {The effect of the Fluorescence Yield selection on the energy scales of Auger, HiRes and TA},
    pdfkeywords = {air-fluorescence yield, energy deposition, Auger, Hi-Res, TA, fluorescence telescopes, ultra-high-energy cosmic rays},
    pdfauthor = {J. R. Vazquez, J. Rosado, D. Garcia-Pinto and F. Arqueros},
    pdfstartview = {FitH},
    }
\begin{document}
\begin{titlepage}
	
\begin{center}
\Huge {\bf The effect of the Fluorescence Yield selection on the energy scales of Auger, HiRes and TA}
\par
\vspace*{1.5cm} \normalsize {\bf {J.~R.~V\'azquez, J.~Rosado, D.~Garc\'{\i}a-Pinto and F.~Arqueros}}
\par
\vspace*{0.5cm} \small \emph{Departamento de F\'{i}sica At\'{o}mica, Molecular y Nuclear, Facultad de Ciencias
F\'{i}sicas, Universidad Complutense de Madrid, E-28040 Madrid, Spain}
\end{center}
\vspace*{2.0cm}
\begin{abstract}
The fluorescence yield data used for shower reconstruction in the Auger, HiRes and TA experiments are
different, not only in the overall absolute value but also in the wavelength spectrum and the various atmospheric
dependencies. The effect on the energy reconstruction of using different fluorescence yield parameterizations is
discussed. In addition, the impact of a change in the fluorescence spectrum depends on the optical efficiency of the
telescopes. A simple analytical procedure allows us to evaluate the combined effect of fluorescence yield and optical
efficiency showing a non-negligible deviation between the energy scales of TA and Auger. However no relevant effect is
found in the comparison between HiRes and Auger. Finally we show that a similar procedure could also be applied with
real data.
\end{abstract}
\end{titlepage}
\section{Introduction}
\label{intro} The accurate measurement of the calorimetric energy of UHECR showers by means of fluorescence telescopes
relies on the precise knowledge of the air-fluorescence yield. In practice, the fluorescence yield is described by a
set of parameters that allows us to convert the fluorescence intensity recorded by the telescope into energy deposited
in the atmosphere at given atmospheric conditions. Auger, HiRes and Telescope Array (TA) use different
parameterizations of the fluorescence yield (absolute value, wavelength spectrum and atmospheric dependency). The
comparison of the energy spectra of Auger and HiRes/TA indicates a possible disagreement in the energy scale
\cite{UHECR2012_WG}, which could be partly due to using different fluorescence yield data. In this work we evaluate the
effect of the fluorescence yield choice on the relative energy scale of these experiments.

While deviations in the absolute value of the fluorescence yield translate directly into deviations in the
reconstructed energy, the effect of differences in the relative wavelength spectra depends on the optical efficiency of
the telescope. Therefore, in order to evaluate the actual influence of the fluorescence yield on the energy scales of
these experiments, the combined effect of the assumed $Y_{\lambda}$ values (i.e., absolute spectral values) and the
optical efficiency $\varepsilon_{\lambda}$ has to be analysed.

The dependence of the fluorescence yield with the atmospheric parameters is determined by the characteristic pressures
$P_{\lambda}'$. Neglecting the humidity contribution to quenching and the temperature dependence of the collisional
cross section, the differences in the assumed $P_{\lambda}'$ values do not lead to significant disagreements in the
energy reconstruction, since most of the shower energy is deposited in the lower atmospheric layers where quenching
effects saturate. However these humidity and temperature contributions to $P_{\lambda}'$ give rise to a non-negligible
effect on the reconstructed energy \cite{astroparticle_monasor}.

The wavelength spectrum assumed by Auger, HiRes and TA, as well as the corresponding optical efficiencies of these
experiments are described in section \ref{FY_datasets}. The analytical method we used to compute the differences in the
energy scales is described in section \ref{sec:Method} and the corresponding results are presented in section
\ref{results}. This theoretical method overestimate somewhat the energy deviations with respect to those from a full
reconstruction of real showers. In section \ref{sec:Method_RD} we discuss on the applicability of this procedure to
real data.

\section{Fluorescence Yield datasets and optical efficiencies}
\label{FY_datasets} 
\begin{figure}%[htb]
\centering
\includegraphics[width=1.0\textwidth]{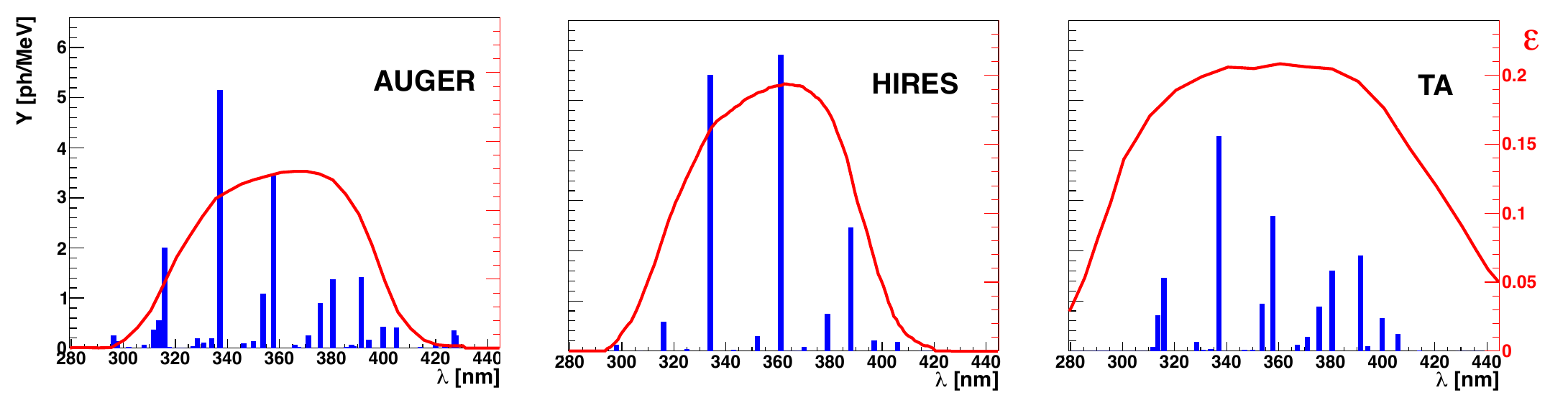}
\caption{\footnotesize{Absolute fluorescence yield (dry air) spectra at 1013 hPa and 293 K employed by the Auger, HiRes and TA
experiments (blue bars). The red line represents the optical efficiency of the corresponding telescopes including
all optical elements}}
\label{fig:yields}
\end{figure}

Data on the fluorescence yield used by Auger, HiRes and TA and the optical efficiency of the
corresponding telescopes are decried below. These parameters are shown together in Figure \ref{fig:yields} for
comparison.

\subsection{Auger}
The fluorescence yield model presently used by the  Pierre Auger Collaboration is defined as follows. The absolute
value of the yield for the 337 nm band $Y_{337}$ is the one measured by Nagano \etal~\cite{nagano}. The yields for the
remaining wavelengths are distributed according to the relative intensities measured by the AIRFLY
Collaboration~\cite{Ave:2007xh}. The characteristic pressures $P_{\lambda}'$ are also those from
AIRFLY~\cite{Ave:2007xh}. Auger is the only experiment taking into account the water vapor quenching of the atmosphere
and the temperature dependence of the collisional cross section \cite{temp_cross_sec}.

The relative optical efficiency of the Auger telescopes \cite{opt_eff_Auger} is dominated by the filter (MUG6 glass
manufactured by Schott-Desag) although other elements (i.e., mirror, corrector ring, and PMT) also contribute.

\subsection{HiRes - TA (Middle Drum)}
The HiRes Collaboration uses the absolute value of the integrated  fluorescence yield (300 - 400 nm) from Kakimoto
\etal~\cite{kakimoto}. The relative intensities of the 337, 351 and 391 nm bands are those measured by Kakimoto while
the intensities of the remaining spectral components are distributed according to those from
Bunner~\cite{bunner_thesis}. The $P_{\lambda}'$ values are those reported by Kakimoto~\cite{kakimoto}.

The optical efficiency of the HiRes telescopes is described in~\cite{tareq_thesis}. The filter is very similar to the
one used by Auger. One of the fluorescence detectors of HiRes has been refurbished and is presently used by the
Telescope Array Collaboration, i.e., the so-called Middle Drum (MD) telescope.

\subsection{Telescope Array (Black Rock Mesa \& Long Ridge)}
The Black Rock Mesa (BRM) and Long Ridge (LR) eyes of the TA Collaboration use the absolute value of the integrated
fluorescence yield (300 - 400 nm ) from Kakimoto et al.~\cite{kakimoto}, that is, the same absolute reference of HiRes.
However TA uses the wavelength spectrum measured by FLASH~\cite{FLASH}. This spectrum has two additional bands not
reported by Kakimoto that are added by TA to get their total absolute yield. The $P'_{\lambda}$ values are also those
from Kakimoto et al.~\cite{kakimoto}.

These BRM and LR telescopes use a BG3 glass (manufactured by Schott) as UV filter. This filter has a higher
transmittance over a wider wavelength interval than the MUG-6 glass employed by Auger. This is a relevant feature that
potentially leads to a lower signal-to-noise ratio. In addition, other molecular bands, not taken into account because
they are absorbed by other narrower UV filters, can contribute to the total fluorescence signal recorded by TA.

\section{Method}
\label{sec:Method} A first estimate of the effect of a change in the fluorescence yield on the reconstructed energy can
be obtained as follows. Neglecting other contributions (e.g., atmospheric transmission and geometrical factors),
the fluorescence signal %$s_i$
recorded by a telescope due to a {\sl true} energy deposition $E_t$ is proportional to $E_t\sum_{\lambda}{Y_{\lambda
t}\varepsilon_{\lambda}}$, where $\varepsilon_{\lambda}$ is the  optical efficiency of the telescope and $Y_{\lambda
t}$ is the {\sl true} fluorescence yield, both dependent on wavelength $\lambda$. Hereafter we relax the notation using
${\bf Y \boldsymbol \varepsilon} \equiv \sum_{\lambda}{Y_{\lambda}\varepsilon_{\lambda}}$.

Let us assume two experiments with efficiencies ${\boldsymbol \varepsilon_1}$ and ${\boldsymbol \varepsilon_2}$ using
different fluorescence yields ${\bf Y_1}$ and ${\bf Y_2}$ respectively. For a given $E_t$ value, the first experiment
reconstructs an energy $E_1$. If the fluorescence yield is replaced by ${\bf Y_2}$, the reconstructed energy would be
$E_{21}$, that is,

\begin{equation}
	\label{eq:ef_1}
	E_{t} {\bf Y_t \boldsymbol \varepsilon_1} = E_1 {\bf Y_1 \boldsymbol \varepsilon_1}=E_{21} {\bf Y_2 \boldsymbol
	\varepsilon_1} \, .	
\end{equation}
Therefore the effect of a change in the fluorescence yield is given by

\begin{equation}
	\label{eq:comp_1}
	\frac{E_{21}}{E_1} = \frac{{\bf Y_1 \boldsymbol \varepsilon_1}}{{\bf Y_2 \boldsymbol \varepsilon_1}} \, .	
\end{equation}
A relationship similar to (\ref{eq:ef_1}) applies to experiment 2

\begin{equation}
	\label{eq:ef_2}
	E_{t} {\bf Y_t \boldsymbol \varepsilon_2} = E_2 {\bf Y_2 \boldsymbol \varepsilon_2}\,.	
\end{equation}

In principle the ratio of the energies measured by these two telescopes $E_2/E_1$ cannot be extracted from the above
expressions. Nevertheless if the {\sl true} fluorescence yield is known, expressions (\ref{eq:ef_1}) and
(\ref{eq:ef_2}) would lead to the solution

\begin{equation}
	\label{eq:comp_true}
	\frac{E_2}{E_1} = \frac{\bf {Y_1 \boldsymbol \varepsilon_1}}{\bf{Y_2 \boldsymbol \varepsilon_2}} \frac{\bf{Y_t
	\boldsymbol \varepsilon_2}}{\bf{Y_t \boldsymbol \varepsilon_1}}\,,	
\end{equation}
and thus, in case one of the two experiments is assumed to use the {\sl true} fluorescence yield (e.g., $\bf {Y_1} =
\bf {Y_t}$), equation (\ref{eq:comp_true}) will give the simple result

\begin{equation}
	\label{eq:comp_21}
	\frac{E_2}{E_1} = \frac{\bf{Y_1 \boldsymbol \varepsilon_2}}{\bf{Y_2 \boldsymbol \varepsilon_2}}\,.
\end{equation}
This is similar to equation (\ref{eq:comp_1}), although for the combined effect of efficiency and yield the ratio has
to be computed with the efficiency of the experiment using the {\sl wrong} yield. Note that even if the assumed {\sl
true} yield turns out to be wrong by a scale factor, expressions (\ref{eq:comp_true}) and (\ref{eq:comp_21}) are still
valid, that is $\bf Y_t$ only needs to describe the {\sl true} relative fluorescence spectrum.

The above procedure allows the evaluation of the deviation in the energy deposition at a given atmospheric depth due to
a change in the fluorescence yield (\ref{eq:comp_1}) or in both fluorescence yield and optical efficiency
(\ref{eq:comp_true}, \ref{eq:comp_21}). The corresponding effect on the total calorimetric energy of the shower can
also be calculated from the above expressions as long as we replace the above dentition, ${\bf Y \boldsymbol
\varepsilon} \equiv \sum_{\lambda}{Y_{\lambda}\varepsilon_{\lambda}}$,  by the following one

\begin{equation}
	\label{eq:def_Ye}
{\bf Y \boldsymbol \varepsilon} \equiv \int_X {\rm d}X \frac {{\rm d}n}{{\rm d}X} G(X) \sum_{\lambda}{Y_{\lambda}(X) \varepsilon_{\lambda}} T_{\lambda}(X)\,,
\end{equation}
where ${{\rm d}n}/{{\rm d}X}$ is the normalised development of energy deposited in the atmosphere by the shower, $G$ is
the geometrical factor, $X$ is the slant depth and $T_{\lambda}$ is the atmospheric transmission of the light
traversing the path from the particular shower point to the telescope location. Using (\ref{eq:def_Ye}) has allowed us
to calculate more realistically the energy deviations. For the development of deposited energy, a Gaisser-Hillas
profile has been used and the $T_{\lambda}$ values have been computed from \cite{rayleigh} (see
\cite{astroparticle_monasor} for details).

\begin{table}[htb]
\caption{Energy deviation with respect to Auger $\Delta E$, defined in (\ref{eq:deltaE}), for different optical
efficiencies (rows) and fluorescence yields (columns). Values (black) in the upper row represent the energy deviation
if Auger would have used the fluorescence model of TA or HiRes. Assuming that Auger uses the {\sl true} fluorescence
spectrum, the energy scale of HiRes (TA) is shifted with respect to that of Auger by the blue (red) value. Results are
shown for three cases: fixed pressure and temperature, realistic Gaisser-Hillas shower without and with temperature and
humidity effects} \label{table1}
  \centering
   \begin{tabular}{||c||c|c||c|c||c|c||}
    \hhline{~*{6}{-}}
    \multicolumn{1}{c||}{} & \multicolumn{2}{c||}{800 hPa} & \multicolumn{2}{c||}{$10^{19}$eV proton} & \multicolumn{2}{c||}{$10^{19}$eV proton} \\
    \multicolumn{1}{c||}{} & \multicolumn{2}{c||}{293 K}   & \multicolumn{2}{c||}{w/o T-h  } & \multicolumn{2}{c||}{T-h} \\
    \hhline{*{7}{-}}
    \backslashbox{$\varepsilon_\lambda$}{$Y_\lambda$} & HiRes & TA & HiRes & TA & HiRes & TA\\
    \hhline{#=#*{3}{=|=#}}
	Auger & 	5 	& 	12 	& 	3   & 	9 	& 	$\sim 0$ 	& 	6 \\
    \hhline{*{7}{-}}
	HiRes & {\textbf{\color{blue}5}} &       & {\textbf{\color{blue}3}} &     & {\textbf{\color{blue} $\sim 0$}} &  \\
    \hhline{*{7}{-}}
    TA    &      & {\textbf{\color{red}16}} &      & {\textbf{\color{red}13}} &   & {\textbf{\color{red}10}} \\
    \hhline{*{7}{-}}
  \end{tabular}
\end{table}

\section{Results}
\label{results} In the first place we present the results neglecting Rayleigh scattering and geometrical factors. Using
the data of section 2, we have obtained the ${\bf Y \boldsymbol \varepsilon}$ values for all the yield-efficiency
combinations. The percentage energy deviations with respect to that of Auger, defined in this work as

\begin{equation}
	\label{eq:deltaE}
	\Delta E = 200 \times \frac{E - E^{\rm Auger}}{E + E^{\rm Auger}}\,,
\end{equation}
have been evaluated at fixed atmospheric conditions (800 hPa and 293 K) for both HiRes and TA (see table 1). The
results in the first row have been obtained using the optical efficiency of Auger and therefore they represent the
energy effect if the Auger yield were replaced by those of the other experiments (\ref{eq:comp_1}). Those of second and
third rows have been evaluated using the efficiencies of HiRes and TA respectively and therefore they give us the
relative energy scale of these experiments with respect to Auger if we assume that Auger is using the {\sl true}
fluorescence spectrum (\ref{eq:comp_21}).

\begin{figure}[h]
%\centering
\includegraphics[width=1.0\textwidth]{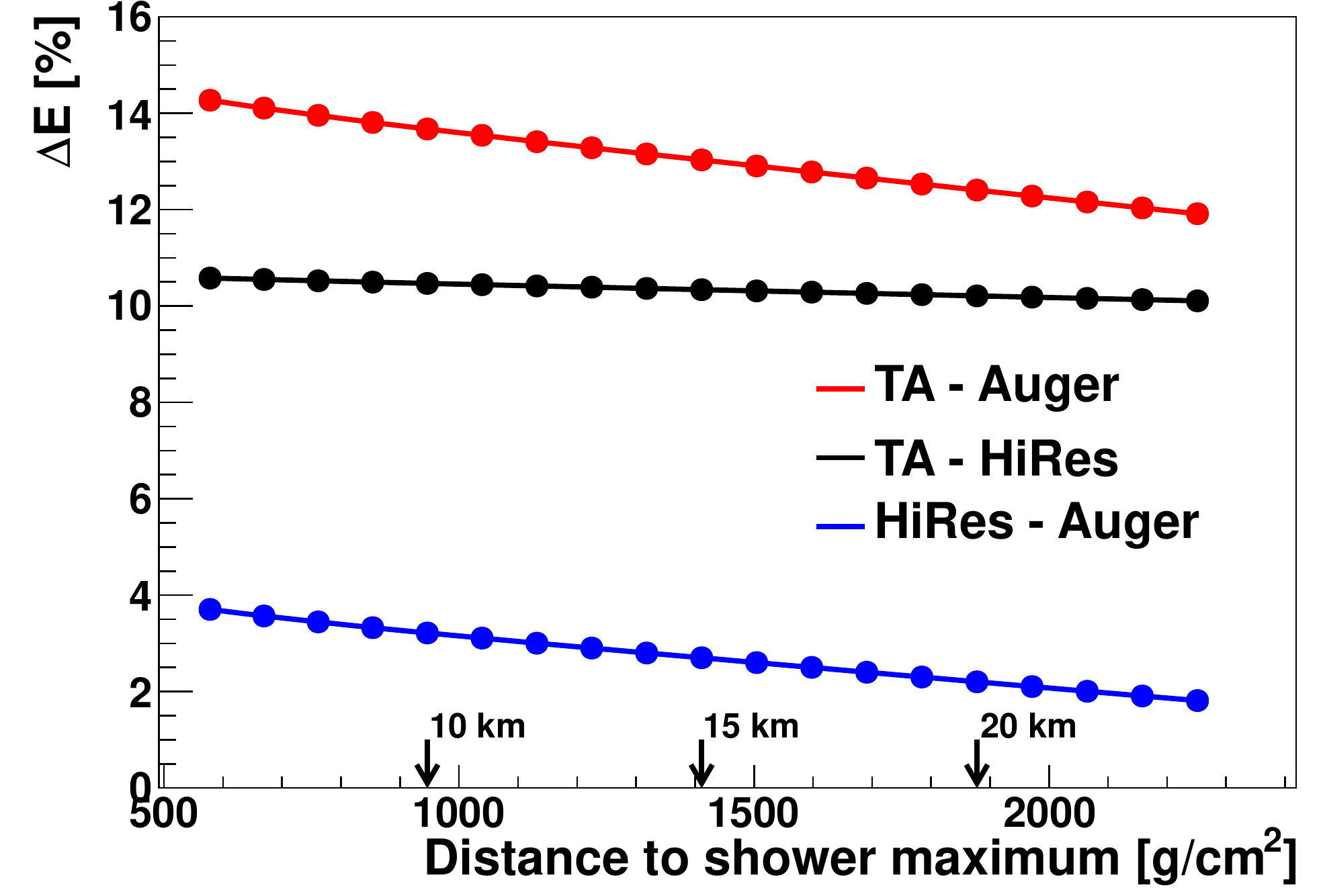}
\caption{\footnotesize{Relative deviations of the energy scales as a function of the distance between the telescope and the
shower maximum for a vertical proton shower of $10^{19}$ eV}. Rayleigh scattering has been included in this calculation.}
\label{fig:Analytical}
\end{figure}

Next a more realistic calculation has been carried out by averaging the energy ratios on the shower development
including the Rayleigh effect and geometrical factors, that is, we have used the ${\bf Y \boldsymbol \varepsilon}$
definition given in expression (\ref{eq:def_Ye}). Calculations for a vertical 10$^{19}$eV proton shower as a function
of the  distance between the telescope and the shower maximum in the range 600 - 2600 ${\rm g}\,{\rm cm}^{-2}$ have
been performed. The results for the above yield-efficiency combinations are plotted in figure \ref{fig:Analytical}. As
shown in the figure, $\Delta E$ depends on the mass thickness between the telescope and the shower maximum due to the
Rayleigh effect. The corresponding $\Delta E$ values for a typical vertical shower at 15km from the telescope are also
shown in the table (columns 3 and 4).

These calculations have been repeated including the temperature and humidity contributions to the $P'_\lambda$ values
in Auger with typical atmospheric profiles (see \cite{astroparticle_monasor} for details). As shown in the table
(columns 5 and 6), the disagreement between the relative energy scales for both cases HiRes-Auger and TA-Auger is
partly cancelled. This is expected since the reconstructed energy increases when the temperature and humidity is
included.

In summary, assuming that the AIRFLY fluorescence spectrum is the {\sl true} one, according to our calculations the
relative energy scale of TA is 13\% (10\%) larger than that of Auger if the temperature and humidity effects are (not)
included in the Auger yield. On the other hand, the Auger energy would increase by 10\% (6\%) if this experiment
replaced its yield by that of TA, without (with) including temperature and humidity effects. Finally we infer that the
energy scale of TA is shifted by about 10\% with respect to that of HiRes.

\section{Towards a precise determination of the relative energy scales}
\label{sec:Method_RD} The above results on the relative energy scales of Auger, HiRes and TA have been obtained
assuming ideal showers that only emit UV radiation from atmospheric fluorescence. In practice a non-negligible amount
of Cherenkov light is emitted and partly detected by the telescope. The contribution of the Cherenkov light has to be
included in the reconstruction algorithm to measure the shower energy \cite{unger_NIM}. The larger the Cherenkov
contribution, the smaller the effect of a change in the fluorescence yield on the reconstructed calorimetric energy. As
a consequence, a given change in the fluorescence yield does not translate entirely in the energy scale and therefore
the above results are expected to overestimate the energy effect in real data.

The effect of a change in the fluorescence yield can easily be  obtained from the observed deviation in the
reconstructed energy of real showers when replacing in the reconstruction algorithm the $Y_{\lambda}$ data by that of
other experiment. However, the evaluation of the combined effect of optical efficiency and fluorescence yield is not
trivial since replacing the optical efficiency in the reconstruction algorithm by that of a different experiment gives
rise to a wrong reconstructed energy. However, as shown next, the problem can be solved following arguments similar to
those presented in section \ref{sec:Method}.

The integrated signals (in appropriate units) recorded by experiments 1 and 2 from a given energy deposition in the
atmosphere will be given by

\begin{equation}
	\label{eq:r_ef_1}
	S_1 = E_{t} {\bf Y_t \boldsymbol \varepsilon_1} = E_1 {\bf Y_1 \boldsymbol \varepsilon_1};
	\hspace{0.5cm}
	S_2 = E_{t} {\bf Y_t \boldsymbol \varepsilon_2} = E_2 {\bf Y_2 \boldsymbol \varepsilon_2}\,.
\end{equation}

In this case ${\bf Y \boldsymbol \varepsilon}$ includes not only the fluorescence contribution given by equation
(\ref{eq:def_Ye}) but also that of the Cherenkov light. In order to determine the relative energy scale of these
experiments ($E_2/E_1$), the energy of given showers should be reconstructed by each experiment using the respective
signal. Nevertheless, if we assume again that experiment 1 is using the {\sl true} fluorescence yield, the relationship
$S_2 = E_{1} {\bf Y_1 \boldsymbol \varepsilon_2}$ holds, that is, we could even calculate the signal of telescope 2
from that of telescope 1. Instead of that, the $E_2/E_1$ ratio can be obtained directly from expression
(\ref{eq:comp_21}) which also applies in this case. While the $\bf Y \boldsymbol \varepsilon$ quantities were
previously computed analytically for modelled showers (section \ref{sec:Method}), now we can make the reconstruction
algorithm to do it for us using real data, including in this way the Cherenkov contribution.

In general the reconstructed energy is proportional to the reciprocal $\bf Y \boldsymbol \varepsilon$ value, the
telescope signal being the proportionality constant, i.e.,  $E = S/{\bf Y \boldsymbol \varepsilon}$. From the $S_1$
signal, $E^*$ parameters can be obtained by reconstructing the energy assuming in the algorithm the optical efficiency
of experiment 2 either with the fluorescence yield of experiment 1 or that of experiment 2,

\begin{equation}
	\label{eq:E*_1}
	E^*_{1} = \frac{S_1}{\bf Y_1 \boldsymbol \varepsilon_2}; \hspace{0.5cm}
	E^*_{2} = \frac{S_1}{\bf Y_2 \boldsymbol \varepsilon_2}\,.
\end{equation}

Note that $E^*_{1}$ and $E^*_{2}$ are not actual values of energy deposition but those of $E_1$ and $E_2$ scaled by the
common $S_1/S_2$ factor,

\begin{equation}
	\label{eq:E*_2}
	E^*_{1} = \frac{S_1}{S_2}E_1; \hspace{0.5cm} E^*_{2} = \frac{S_1}{S_2}E_2\,,
\end{equation}
and therefore the relative energy scale can be determined from

\begin{equation}
	\label{eq:comprd_21}
	\frac{E_2}{E_1} = \frac{E_2^*}{E_{1}^*}\,.
\end{equation}

From arguments similar to those of section (\ref{sec:Method}), the condition for experiment 1 of using the {\sl true}
fluorescence yield can be relaxed to just using the {\sl true} fluorescence spectrum, although in this case we have to
assume that the Cherenkov contribution is small with respect to that of fluorescence.

\section{Conclusions}
\label{conclusions} In this work we have evaluated the effect of the fluorescence yield choice on the energy
reconstruction of extensive air showers. Using a simple analytical procedure we found that Auger (without T and h
effects) would increase the energy by about 9\% and 3\% if the fluorescence yield of TA and HiRes respectively were
used instead.

For the comparison of the energy scales, the optical efficiency of the experiments also matters. A simple procedure
allows to carry out this comparison analytically. Assuming that Auger uses the {\sl true} fluorescence spectrum, i.e.,
AIRFLY \cite{Ave:2007xh}, we found that the TA energy is larger than that of Auger (including temperature and humidity
effects) by about 10\%. Notice that this energy deviation is accidentally lowered because TA does not take into account
the temperature and humidity contributions to the fluorescence yield. Otherwise the energy scale of TA would be 13\%
larger.

The above results are somewhat overestimated because the analytical procedure shown here does not take into account the
contribution of Cherenkov light in the reconstruction of real data. A procedure has been described that allows to make
these comparisons from the reconstruction of real data of any experiment. Some tests have shown that the analytical
$\Delta E$ values are overestimated with respect to those of real data by about 15\% and therefore the energy scale of
TA would be larger than that of Auger by about 6\% (10\%) if the humidity and temperature effects are (not) taken in
account in Auger.

Finally, from the present analysis we also conclude that the energy of TA (BRM and LR eyes) is about 10\% larger than
that of HiRes (i.e., MD eye of TA) due to the different fluorescence spectra used by these experiments.

\section{Acknowledgement}
This work has been supported by the former Ministerio de Ciencia e Innovaci�n (FPA2009-07772 and CONSOLIDER CPAN
CSD2007-42). Data of the fluorescence yield and optical efficiency of TA have been kindly provided by Y.~Tsunesada and
D.~Ikeda. Fruitful discussions with our colleagues of the Auger collaboration are acknowledged.

\end{document}